\documentclass{iau_FM}
\usepackage{graphicx}

\graphicspath{{figures/}} 

\title[UHECRs and CMFs] 
{Propagation of UHECRs in the local Universe and origin of cosmic magnetic fields}

\author[Stefan Hackstein]   
{S. Hackstein$^1$, F. Vazza$^2$, M. Br\"uggen$^1$, J. G. Sorce$^3$, S. Gottl\"ober$^3$}

\affiliation{$^1$ Sternwarte Bergedurf, Universit\"at Hamburg \\ Gojenbergsweg 112,
21029, Hamburg, Germany \\
$^2$ INAF, Istituto di Radioastronomia di Bologna, via Gobetti 101, I-41029 Bologna, Italy\\
$^3$ Universit\'e de Strasbourg, CNRS, Observatoire astronomique de Strasbourg, UMR 7550, F-67000 Strasbourg, France \\
$^4$ Leibniz Institute for Astrophysics Potsdam, An der Sternwarte 16, 14482 Potsdam
 \\ email: {\tt stefan.hackstein@hs.uni-hamburg.de,

} }

\pubyear{2018}
\jname{Astronomy in Focus} 
\editors{Teresa Lago, ed.}
\begin{document}

\maketitle

\begin{abstract}
We simulate the propagation of cosmic rays at ultra-high energies, $\gtrsim 10^{18}$ eV, in models of extragalactic magnetic fields in constrained simulations of the local Universe. 
We investigate the impact of different magneto-genesis scenarios, both, primordial and astrophysical, on the propagation of cosmic rays.
Our study shows that different scenarios of magneto-genesis do not have a large impact on the anisotropy measurements.
The distribution of nearby sources causes anisotropy at very high energies, independent of the magnetic field model.
We compare our results to the dipole signal measured by the Pierre Auger Observatory.
All our models could reproduce the observed dipole amplitude with a pure iron injection composition.
This is due to clustering of secondary nuclei in direction of nearby sources of heavy nuclei.
 A light injection composition is disfavoured by the non-observation of anisotropy at energies of 4 - 8 EeV.

\keywords{MHD, methods: numerical, ISM: magnetic fields, cosmic rays}
\end{abstract}

\firstsection 
\section{Introduction}
We investigate constrained Magnetohydrodynamical (MHD) simulations of the local Universe for different scenarios of magneto-genesis and probe their influence on the propagation of ultra-high energy cosmic rays (UHECRs).
We compare different source models to see what information can be inferred on the extragalactic magnetic field.
The MHD simulations used constrained initial conditions \cite[(Sorce et al. 2016)]{Jenny} and were performed with the MHD code ENZO\footnote{http://www.enzo-project.org}, which is an AMR code for cosmological simulations.
The resulting magnetic fields are used with the CRPropa\footnote{https://crpropa.desy.de/} code to simulate the propagation of UHECRs. 
The observer is placed at the center of the simulation, the defined position of the Milky Way.
\\ 
We investigate two major scenarios for magneto-genesis: magnetic fields of primordial origin - prior to z = 60, where the MHD simulations start - with several spectral indices of the initial magnetic power spectrum (more info in \cite[Hackstein et al. 2018]{Hackstein18}).
We compare this scenario to an astrophysical origin, modelled with thermal and magnetic feedback of AGN at several different redshift ranges.
We improve on \cite[Dolag et al. 2003]{Dolag} by testing these different scenarios and by obtaining the expected $B$-$\rho$-ratio from a single consistent MHD simulation.
Both scenarios reproduce magnetic fields observed in clusters reasonably well. 
The main difference is that in the low-density regions, i. e. the voids, the magnetic field strength differs by up to 10
orders of magnitude.
The two scenarios are therefore well suited to highlight the influence of the low-density regions.

\section{Results}

\begin{figure}[t]
\includegraphics[width=9.6cm]{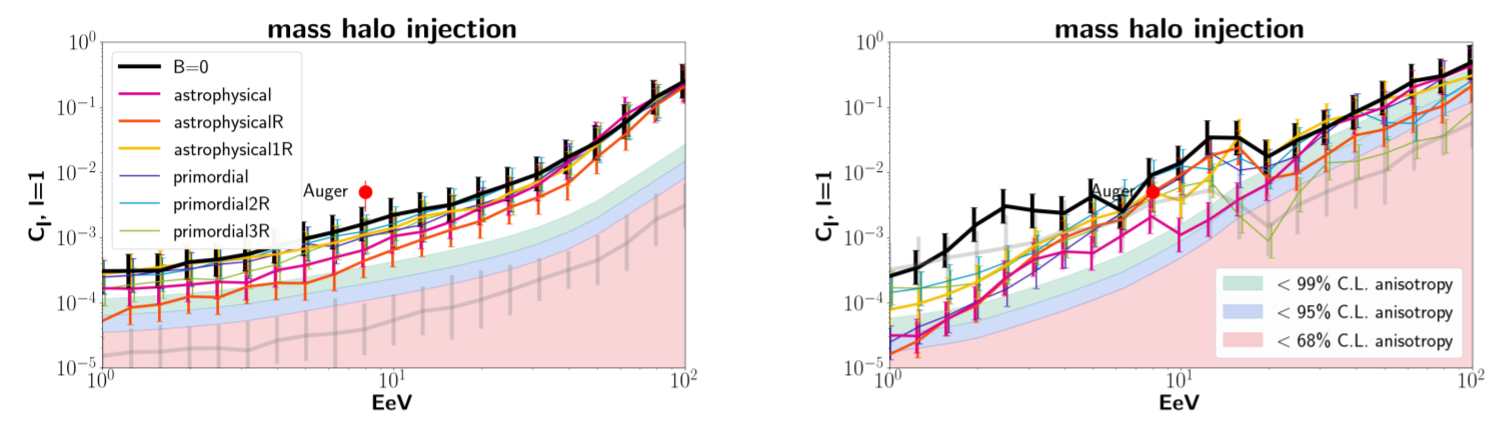}
\includegraphics[width=3.4cm]{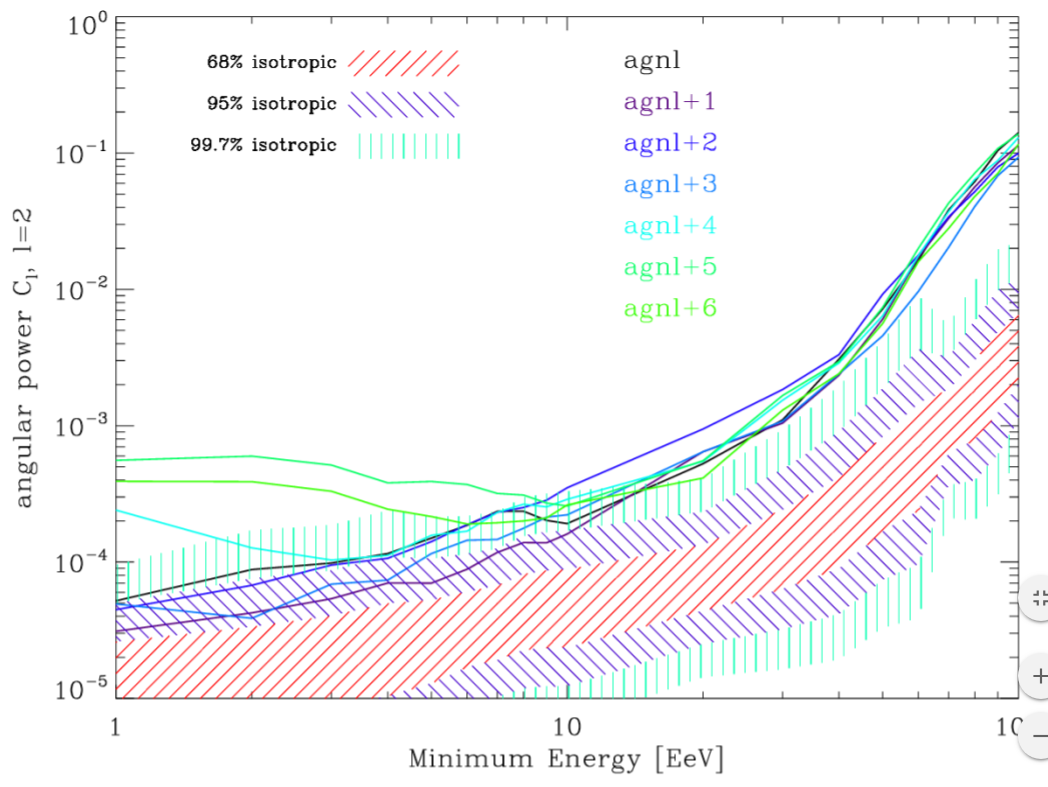}
\caption{ Results for angular power as function of minimum energy of considered events.
The coloured bands indicate the confidence level of anisotropy.
The dipole observed in the Auger data $> 8$ EeV is indicated.
Left: Dipole from pure proton injection. Center: dipole from pure iron injection. Right: Quadrupole from pure proton injection }
\label{fig2}
\end{figure}
Anisotropy is measured with the angular power spectrum of spherical decomposition of the full sky of observed UHECR events.
We consider as sources the center of viral halos identified in the MHD simulation, where the most energetic objects are most likely to be found. 
This source density is at the lower bound of allowed densities, $\gtrsim 10^{−4} \rm\ Mpc^{−3}$, as shown by \cite[di Matteo \& Tinyakov 2017]{Matteo}.
\\
Results for the different scenarios of magneto-genesis are pretty converged, the fields in voids have minor influence on the large scale anisotropy signal of UHECRs. 
It is unlikely that we can identify magneto-genesis scenarios by full-sky observations of UHECR events without prior knowledge on their sources.
For light injection composition, in order to reproduce the Auger dipole, a rather extreme source model is needed, that might already be at odds with present constraints.
For a heavy composition the contribution of the most nearby sources increases, due to multiple allowed hits of secondary nuclei of the same primary particle.
This effect is the strongest at $E = E_{\rm max}/A$, the maximum energy of secondary protons that can reach the observer only from the most nearby sources.
\\ 
In another set of simulations, anisotropy occurs at low energies for an increased magnetic field strength, independent of the distribution of sources (results from \cite[Hackstein et al 2016]{Hackstein16}).
Here, a uniform field dominates around the sources. 
One can show that anisotropy is indeed expected in that case.
The same is true for many fields with pronounced component in vertical direction.
This offers the chance to put limits on such fields in the halo of the Milky Way.

\section{Conclusions}
\begin{enumerate}
\item  light injection composition is unlikely to reproduce the Auger dipole while heavy injection increases multipoles.
\item anisotropy at highest energies enables UHECR astronomy
\item magneto-genesis scenario unlikely to be distinguished by UHECR observations
\item strong vertical magnetic field component introduces quadrupole anisotropy $\Rightarrow$ limit magnetic fields in Milky Way halo.

\end{enumerate}


\begin{thebibliography}{}

\bibitem[Matteo]{Matteo}
{di Matteo, A. and Tinyakov, P.} 2018, 
\textit{MNRAS}, 476, 715

\bibitem[Dolag]{Dolag}
{Dolag K., Grasso D., Springel V., Tkachev I.} 2004, 
\textit{JETP Lett.}, 79, 583

\bibitem[Hackstein16]{Croat_etal05}
{Hackstein, S., Vazza, F., Br\"uggen M., Sigl, G., Dundovic, A.} 2016, 
\textit{MNRAS}, 462, 3660

\bibitem[Hackstein18]{Busso_etal99}
{Hackstein, S., Vazza, F., Br\"uggen, M., Sorce, J. G., Gottl\"ober, S.} 2018, 
\textit{MNRAS}, 475, 2519

\bibitem[Jenny]{Jenny}
{Sorce, J. G. et al.} 2016,
\textit{MNRAS}, 455, 2078


\end{thebibliography}
\end{document}